\documentstyle[sprocl,epsf]{article}

\def\be{\begin{equation}}
\def\ee{\end{equation}}
\def\bea{\begin{eqnarray}}
\def\eea{\end{eqnarray}}

\begin{document}

\title{The Mass of the $b$ Quark from Lattice NRQCD}

\author{S.~Collins}


\address{The Department of Physics and Astronomy, The University of Glasgow,\\ Scotland, G12 8QQ} 


\maketitle\abstracts{We present results for the mass of the $b$ quark in the
$\overline{MS}$ scheme obtained by calculating the binding energy of
the $B$ meson in the static limit. The self energy of a static quark,
$E_0^\infty$ needed for this purpose, is now known to $O(\alpha^3)$ in
the quenched approximation. We find a {\it preliminary} value of
$\overline{m_b}(\overline{m_b})=4.34(7)$~GeV at $n_f=0$. The error is
dominated by the remaining uncertainty in $E_0^\infty$.  In addition,
using $E_0^\infty$ at $O(\alpha^2)$, we estimate that the quark mass
is reduced by approximately $70$~MeV when two flavours of dynamical
quarks are introduced.}

\section{Introduction}

The mass of the $b$ quark in the $\overline{MS}$
scheme~($\overline{m_b}$) can be extracted on the lattice using NRQCD,
via the pole mass, $M_{pole}$, by calculating the binding energy of
the $B$ meson, $\overline{\Lambda}_{bind}$: $M_{pole} = M^{expt}_B -
\overline{\Lambda}_{bind}$ where $M^{expt}_B$ is the spin-average of
the experimental $B$ and $B^*$ masses and $\overline{\Lambda}_{bind}=
E_{sim}-E_0$. $E_{sim}$ is the energy of the $B$ meson~(at rest) in
NRQCD and $E_0$ is the $b$ quark self energy. The pole mass is then
converted to $\overline{m_b}$ at some scale $\mu$ using the continuum
perturbative factor $Z_{cont}$: $\overline{m_b}(\mu) = Z_{cont}(\mu)
M_{pole}$.  While $M_{pole}$ has an $O(\Lambda_{QCD})$ renormalon
ambiguity, this is cancelled by similar effects in $Z_{cont}$:
$\overline{m_b}$ is well defined.

At present, $E_0$ is only known to $O(\alpha)$ for the $b$
quark. However, in the limit of the $b$ quark mass becoming infinite,
$E_0$ is known to $O(\alpha^3)$ if internal quark loops are
neglected~(the quenched approximation). The tadpole-improved formula
can be expressed as
\begin{equation}
E_0^\infty = 1.070\alpha_p + 0.118\alpha_p^2 - 0.3(1.4)\alpha_p^3 \hspace{.3cm}:\hspace{.3cm}\alpha_p=\alpha_p^{(n_f=0)}(0.84/a).\label{e0}
\end{equation}
The $\alpha_p^3$ coefficient has been determined by Lepage
et. al.~\cite{trottier}. The error on the coefficient is numerical and
quite large. However, it provides a realistic estimate of the
uncertainty in $E_0$, compared to using $2-loop$ perturbation theory
and assuming the contribution of higher order terms is $1\alpha_p^3$.

In Eq.~\ref{e0}, $E_0$ is expressed in terms of a coupling
constant defined on the lattice from the plaquette,
$\alpha_p$~\cite{alphap}, evaluated at a characteristic gluon momenta,
$q*=0.84/a$, calculated using the BLM proceedure~\cite{tadpole,qstar}. In
addition, the lattice calculation of $E_{sim}$ has been tadpole
improved, whereby all gauge fields on the lattice are divided by a
`mean-field' approximation to the gluon field, $u_0$, to obtain more
continuum-like operators. The corresponding tadpole improvement of
$E_0$ leads to the addition of the perturbative series for $\ln
u_0$. These ingredients result in a well behaved perturbative series
for $E_0$. This is certainly not the case if the bare lattice
coupling, $\alpha_L=g_0^2/(4\pi)$ is used:
\begin{equation}
E_0^\infty = 2.1173\alpha_L + 11.152\alpha_L^2 + 82.7(1.4)\alpha_L^3
\end{equation}
Di~Renzo et. al.~\cite{renzo} have also determined the $O(\alpha^3)$
coefficient. They obtain $86.2(.6)$, when $\alpha_L$ is
used. Encouragingly, the two determinations, which have very different
systematic errors, agree within $3\sigma$.

$Z_{cont}$ has been calculated to $\alpha_p^3$ 
by Melnikov and van~Ritbergen~\cite{zcont}: 
\begin{equation}
Z_{cont} = 1 - 0.4244\alpha_p - 0.4771\alpha_p^2 - 1.814\alpha_p^3 \hspace{.3cm}:\hspace{.3cm}\alpha_p=\alpha_p^{(0)}(0.62\overline{m_b}).
\end{equation}
The series is well-behaved and we estimate the
uncertainty in $Z_{cont}$ to be $3\alpha_p^4$.

The error introduced by working in the static limit, i.e. ignoring
${\small O(\Lambda_{QCD}/M)}$ contributions to $\bar{\Lambda}_{bind}$,
leads to $\approx 1\%$ uncertainty in $\overline{m_b}$. The error
arising from working in the quenched approximation is also likely to
be around $1\%$~(assuming a $10-20\%$ shift in $E_{sim}$ when sea
quarks are included).  These effects are the same size as the error
arising from the numerical error in $E_0$.

We obtained $E_{sim}^\infty$ by extrapolating the simulation energy
calculated at finite heavy quark mass. The latter was obtained as part
of a high statistics study of the $B$ meson spectrum in the quenched
approximation at three lattice spacings~($a$), with
$a^{-1}=1{-}2.5$~GeV. For details of the simulations see
reference~\cite{joachim}. Note that we use the spin-average of the
experimental masses for the $B$ and $B^*$ mesons in the expression for
$M_{pole}$ in order to reduce the error in using $E_{sim}^\infty$. In
addition, we performed a study of sea quark effects using results
obtained from a simulation including two flavours of sea
quarks~($n_f=2$)~\cite{me} with $a^{-1}\sim 2$~GeV.  Only the
$O(\alpha_p^2)$ coefficient for $n_f=2$ has been computed~\cite{junko}
and hence the comparison with $\overline{m_b}$ at $n_f=0$ is performed
using $E_0$ and $Z_{cont}$ to this order.

\section{Results}

Tables~\ref{quench} and~\ref{loop} summarize our results.  Within the combined
statistical and systematic errors we see that this is the case and we
take the result at $\beta=6.0$ as our best determination of
$\overline{m_b}(\overline{m_b})=4.34(7)$~GeV at $n_f=0$.
Note that the numerical error in $E_0$ dominates the
uncertainty in $\overline{m_b}$.

In addition, using the results at $\beta=6.0$ at $n_f=0$ from the
$B_s$ meson and those obtained at $n_f=2$ we see that the $b$ quark
mass decreases by $70$~MeV at $O(\alpha_p^2)$ when sea quarks are
introduced. Assuming the systematic~(perturbative) errors for the two
simulations are correlated this is $\sim 2\sigma$ in the~(remaining)
statistical errors and the same size as the error in
$\overline{m_b}(\overline{m_b})$ at $O(\alpha_p^3)$. Further work is
necessary to reduce the error in $E_0$, in order for sea quark and
$O(\Lambda_{QCD}/M)$ effects to be significant.

\begin{table}
\begin{center}
\caption{$\overline{\Lambda}_{bind}^\infty$ and $\overline{m_b}(\overline{m_b})$ in GeV from the $B$ meson at $n_f=0$. The statistical and main systematic errors are estimated, including those
due to determining the inverse lattice spacing~($a^{-1}$) and residual
discretisation effects in $E_{sim}$ $\sim
O((\Lambda_{QCD}a)^2)$.\label{quench}}
\vspace{-0.25cm}
\begin{tabular}{|cccccc|cccccc|}\hline
 & \multicolumn{5}{c|}{$\overline{\Lambda}_{bind}^\infty$} &
 \multicolumn{6}{c|}{ $\overline{m_b}(\overline{m_b})$} \\\hline
 $\beta$ & & stat. & $E_0^\infty$ & $a^{-1}$ & disc. & & $Z_{cont}$ &
 stat & $E_0^\infty$ & $a^{-1}$ & disc. \\ & & & $1.4\alpha_p^3$ & & &
 & $3\alpha_p^4$ & &$1.4\alpha_p^3$ & & \\\hline 5.7 & .24 & (1) &
 (11) & (1) & (9) & 4.43 & (3) & (1) & (10) & (1) & (8) \\ 6.0 & .35 &
 (2) & (6) & (1) & (4) & 4.34 & (3) & (2) & (5) & (1) & (3) \\ 6.2 &
 .36 & (8) & (5) & (2) & (2) & 4.32 & (3) & (7) & (4) & (2) & (2)
 \\\hline
\end{tabular}
\end{center}
\end{table}

\begin{table}
\begin{center}
\caption{The change in $\overline{m_b}(\overline{m_b})$ from $n_f=0$ to $2$, where 2-loop perturbation theory is used.\label{loop}}
\begin{tabular}{|ccc|}\hline
 $n_f$ &
 $\overline{m_b}(\overline{m_b})$ & stat. error \\\hline 
 0 & 4.45 & .01 \\ 2 & 4.38 & .02 \\\hline
\end{tabular}
\end{center}
\end{table}

\section*{Acknowledgements}
The author acknowledges support from the Royal Society of Edinburgh.

\end{document}